\documentclass[runningheads]{llncs}

\usepackage[T1]{fontenc}
\usepackage{graphicx}

\usepackage{tikz} %
\usepackage{pgfplots}
\usepackage{tabularx}
\usepackage{booktabs}
\usepackage{csquotes}
\usepackage{paralist}

\usepackage{graphicx}
\usepackage{hhline} %
\usepackage{url}
\usepackage{mathtools}
\usepackage{cleveref}
\usepackage{xspace}

\newcommand*{\ie}{i.e.\@\xspace}
\usepackage{balance}

\usepackage{comment}
\usepackage{framed}
\usepackage{textpos}

\usetikzlibrary{positioning}
\def\BibTeX{{\rm B\kern-.05em{\sc i\kern-.025em b}\kern-.08em
    T\kern-.1667em\lower.7ex\hbox{E}\kern-.125emX}}

\definecolor{node_blue}{RGB}{51, 153, 255}
\definecolor{node_red}{RGB}{230, 0, 0}
\definecolor{node_green}{RGB}{51, 204, 51}

\usetikzlibrary{shapes.geometric, arrows, positioning, calc}

\usepackage{makecell}

\tikzset{
    every node/.style={font=\small},
    block/.style={rectangle, draw, text centered, rounded corners, minimum height=1cm},
    flink/.style={rectangle, draw=green!50, fill=green!20, thick, minimum height=1cm},
    prink/.style={rectangle, draw=blue!50, fill=blue!20, thick, minimum height=1cm},
    dashedblock/.style={rectangle, draw=green!50, fill=green!20, thick, dashed, minimum width=2cm, minimum height=1cm},
    interface/.style={rectangle, draw=blue!50, thick, minimum width=3cm, minimum height=1cm, text centered},
    arrow/.style={-stealth, thick},
    doublearrow/.style={<->, thick},
    dashedarrow/.style={-stealth, thick, dashed},
}

\begin{document}

\title{Prink: $k_s$-Anonymization for Streaming Data in Apache Flink}
\titlerunning{Prink: $k_s$-Anonymization for Streaming Data in Apache Flink}

\author{
Philip Groneberg\inst{1} \and
Saskia Nuñez von Voigt\inst{1} \orcidID{0009-0001-2163-8359} \and\\
Thomas Janke\inst{1} \orcidID{0009-0009-6021-9817} \and
Louis Loechel\inst{1} \orcidID{0000-0002-5877-3706} \and\\
Karl Wolf\inst{1} \orcidID{0000-0003-4607-782} \and
Elias Grünewald\inst{2} \orcidID{0000-0001-9076-9240} \and
Frank Pallas\inst{3} \orcidID{0000-0002-5543-0265}
}

\institute{
Technische Universität Berlin, Germany \and
Charité – Universitätsmedizin Berlin, Germany \and
Paris Lodron Universität Salzburg, Austria
}

\authorrunning{P. Groneberg et al.}

\maketitle

\begin{textblock*}{1\textwidth}(0cm,-8.5cm) %
\begin{center}
\begin{framed}
    \textit{Preprint (2025-05-19) before final copy-editing of the accepted peer-reviewed paper to appear in the Proceedings of the\\ \textbf{20\textsuperscript{th} International Conference on\\ Availability, Reliability and Security (ARES'25)}.}
\end{framed}
\end{center}
\end{textblock*}

\begin{abstract}
In this paper, we present Prink, a novel and practically applicable concept and fully implemented prototype for $k_s$-anonymizing data streams in real-world application architectures. Building upon the pre-existing, yet rudimentary CASTLE scheme, Prink for the first time introduces semantics-aware $k_s$-anonymization of non-numerical (such as categorical or hierarchically generalizable) streaming data in a information loss-optimized manner. In addition, it provides native integration into Apache Flink, one of the prevailing frameworks for enterprise-grade stream data processing in numerous application domains.

Our contributions excel the previously established state of the art for the privacy guarantee-providing anonymization of streaming data in that they 1) allow to include non-numerical data in the anonymization process, 2) provide discrete datapoints instead of aggregates, thereby facilitating flexible data use, 3) are applicable in real-world system contexts with minimal integration efforts, and 4) are experimentally proven to raise acceptable performance overheads and information loss in realistic settings. With these characteristics, Prink provides an anonymization approach which is practically feasible for a broad variety of real-world, enterprise-grade stream processing applications and environments.

\end{abstract}

\section{Introduction}

Stream processing is a core paradigm underlying many modern, enterprise-grade application architectures. From smart energy infrastructures over nearly real-time traffic monitoring and optimization to assistive environments, %
data are increasingly processed and used on the fly and in real-time while flowing through stream processing pipelines without ever being permanently persisted at all. At the same time, respective applications often collect and process personal and sometimes highly sensitive data, calling for appropriate anonymization.

A widely used approach for privacy protection is $k$-anonymity~\cite{sweeney2002k}, which ensures
that each entry within a group of $k$ entries is indistinguishable from the others.
Extensions such as $\ell$-diversity~\cite{machanavajjhala2007diversity} and $t$-closeness~\cite{4221659} further mitigate risks related to attribute disclosure. These approaches are particularly prominent in scenarios where detailed individual-level data are not needed, but group-level patterns must be preserved.
Energy district management serves as a prime example where such privacy guarantees are essential. Here, the focus is not on individual household consumption, but on understanding broader usage patterns across neighborhoods or similar households.

An alternative fundamental concept, differential privacy~\cite{dwork2006differential}, provides strong mathematical semantic privacy guarantees
by adding calibrated noise to the data.
While differential privacy provides privacy guarantees beyond the ones of $k$-anonymity, it introduces noise into data, \ie,
specific energy consumption values are inherently uncertain, making stable trend analysis more difficult.
For instance, when forecasting energy demand, fluctuations caused by differential privacy could obscure real consumption patterns, leading to suboptimal infrastructure planning.
In contrast, $k$-anonymity preserves data consistency, ensuring that neighborhood-level energy statistics remain reliable over time, which is crucial for energy management.

Despite its advantages, traditional $k$-anonymity methods were not designed for streaming environments, where data must be anonymized in nearly real time. 
Stream-specific adaptations of $k$-anonymity, such as $k_s$-anonymity~\cite{cao2010castle,castleguard}, present a promising prospect by achieving privacy guarantees comparable to their non-streaming counterparts but often focus on numerical values, limiting their applicability to data sets that also contain categorical or hierarchical attributes.
Additionally, enterprise-grade systems require efficient, low-latency processing to handle large data volumes without significant performance overhead.
These factors are critical for driving the practical adoption of advanced privacy-preserving schemes and for shaping potential regulatory requirements for their implementation in real-world systems \cite{pallas-ea2024_roadmap}.

To overcome these limitations and advance the practical applicability of guarantee-providing anonymization schemes in stream-based application architectures, we make the following contributions: %
\begin{itemize}
    \item We present Prink, a privacy-preserving stream anonymization framework that extends the pre-existing CASTLE algorithm and implementation for $k_s$-anonymization with $\l$-diversity to support categorical and hierarchical generalization, along with the ability to handle multiple sensitive attributes in $l$-diversity.
    \item We propose a novel scheme for semantics-aware information loss scheme in the anonymization of non-numerical streaming data, incorporating the support for dynamic generalization trees.
    \item We introduce a concept for distributing CASTLE's clustering approach across multiple nodes while preserving $k-$ and $l$-guarantees to achieve the scalability required in real-world streaming applications. %
    \item We provide an open-source implementation designed to seamlessly integrate these functionalities into real-world use cases, employing the established and highly scalable stream-processing framework Apache Flink\footnote{\url{https://flink.apache.org/}}.
    \item We conduct an experimental evaluation to assess the performance overheads, focusing on latency in realistic settings, demonstrating the practical viability of our approach.
\end{itemize}

The structure of this paper is as as follows. In \Cref{background}, we introduce relevant background and related work. Our general approach and the details of our stream-specific anonymization scheme for non-numerical data are provided in \Cref{sec:approach}. Our evaluation and results are presented in \Cref{eval} and further discussed in \Cref{discussion}, before we conclude our paper in \Cref{conclusion}.

\section{Background and Related Work}
\label{background}
In the following, we provide relevant preliminaries for advanced anonymization in streaming architectures.

\subsection{Anonymization Techniques and Anonymity Guarantees}

To assure the %
privacy of individuals in the processing of data referring to them, different anonymization techniques are used. 
These techniques mostly include perturbation, generalization or basic data reduction \cite{gruschka2018privacy,majeed2020anonymization,marques2020analysis,ashkouti2021di}, and provide some level of anonymization. To make these levels measurable and more usable, and to avoid unexpected privacy violations resulting from outlier datapoints, anonymity guarantees such as $k$-anonymity \cite{sweeney2002k}, $\ell$-diversity \cite{machanavajjhala2007diversity} or $t$-closeness \cite{4221659} have been established, particularly for publicly releasing static data sets for onward use. %
These provide very specific and exact levels of anonymity, guaranteeing, for instance, that at least $k$ individuals are indistinguishable from each other in the anonymized data set or that each generalized cluster contains at least $\ell$ different values for the sensitive attribute. %

Since all anonymization techniques reduce or alter the original data, some information is necessarily lost during the anonymization process. To not only provide anonymity guarantees, but also maintain the value of shared, anonymized data (e.g., for subsequent analytics), advanced techniques such as range aggregation for numerical data (assigning distinct values to value ranges) and hierarchical generalization \cite{towards_k_anon_non_numerical,MARTINEZ2013294,simi_data_anonymization,samarati1998protecting} for non-numerical data are used instead of, for instance, removing attributes completely. These techniques provide not just the requirements for the mentioned privacy guarantees, but also try to keep the resulting information loss as low as possible.

Having been established in the 2000s, most respective techniques
only work on static data sets such as census data  \cite{sweeney2002k,machanavajjhala2007diversity}. %
Further related work applied such techniques to big data architectures \cite{sopaoglu2017top,suneetha2020novel} including domain hierarchy approaches \cite{bazai2021scalable}.
For settings with dynamic data, such as frequently updated databases or even continuous data streams, however, they cannot be applied without breaking the to-be provided guarantees. Given that such data-intensive settings particularly shape the collection, processing and use of personal data today, alternative approaches are needed.

\subsection{Anonymity Guarantees for Dynamic and Streaming Data}
To provide anonymity guarantees and techniques that minimize information loss for dynamic and streaming data, new solutions needed to be created. Respective approaches can be categorized into those following the notion of differential privacy and those adapting pre-existing concepts from $k$-anonymity (and extensions) to stream-specific givens.

With differential privacy \cite{dwork2006differential,dwork2014algorithmic}, dynamic and streaming data are typically handled through aggregation and noising, ensuring that the impact of individual data points on results remains within specified boundaries.\footnote{Specifically, differential privacy guarantees that the probabilities of outcomes, such as aggregation results, remain nearly identical for two neighboring data sets that differ by only one data point.}
In global differential privacy, a trusted centralized entity holds all raw data
and processes queries by providing noisy results, such as sums, counts, or averages.
The added noise is calibrated based on the sensitivity of a query and the differential privacy parameter $\varepsilon$ that defines the privacy guarantee.
Use cases for global differential privacy span a wide range of application, including statistical databases (e.g., for census data \cite{abowd2018us}), privacy-preserving social network analysis \cite{jiang2021applications}, or  large-scale trip data analysis \cite{bassolas2019hierarchical},
often extending to advanced types of aggregations like graph metrics \cite{kasiviswanathan2013analyzing} or geospatial analyses \cite{bassolas2019hierarchical}.

In local differential privacy~\cite{dwork2014algorithmic,kasiviswanathan2011can}, data are not transmitted in their raw form to a trusted curator but are instead locally anonymized by adding noise before being released.
This local noise addition ensures that individual data entries remain private.
Data structures such as Bloom filters \cite{erlingsson2014rappor} and FM sketches \cite{nunez2020rrtxfm} enable differentially private aggregations, including popularity statistics \cite{cormode2018privacy} and cardinality estimations \cite{nunez2020rrtxfm}.
Local differential privacy is particularly useful for scenarios where many users must report their data to an untrusted party while preserving privacy, such as in location-based services \cite{andres2013geo,kim2021survey}, smart metering \cite{eibl2017differential}, and large-scale telemetry data collection \cite{ding2017collecting}.

Although differential privacy provides strong privacy guarantees, it inherently limits the use of the resulting data—especially in the case of local differential privacy.
This limitation arises because the data can only be processed in ways that were anticipated and planned for before implementing a specific, carefully chosen differential privacy mechanism.
Additionally, achieving reasonable data utility with differential privacy often requires large volumes of data, further restricting its practical applicability.
Moreover, advanced differential privacy mechanisms that extend beyond numerical data are frequently less reusable and necessitate extensive customization for each use case, sometimes introducing unforeseen and non-obvious re-identification risks \cite{houssiau2022difficulty}.

Proposals for adapting established approaches of $k$-anonymity (and extensions) to the specifics of data streams, in turn, initially were 
of theoretical nature and came without \cite{cao2010castle,al2018experimenting} or with only rudimentary prototype implementations \cite{castleguard}. Only recently, research on integrating them into real-world streaming systems has gained momentum \cite{redCastle}. Respective endeavors have, however, so far focused on smaller and non-distributed systems, leaving important streaming frameworks such as Kafka or Apache Flink unsupported. 

Existing concepts and implementations merely focus on generalizing numerical and occasionally categorical data for anonymization purposes \cite{apuan2011landscape}. In reality, however, the data to be anonymized also comprises non-numerical attributes. Different from established %
$k$/$\ell$/$t$-schemes for static data, these are so far not properly covered by existing stream-focused anonymization schemes, especially with regard to the reduction of information loss.

Within these limitations, however, the CASTLE algorithm \cite{cao2010castle} has established as the prevailing approach for $k$-anonymizing data streams over alternative ones such as KIDS \cite{zhang2010kids} or K-VARP \cite{otgonbayar2018k}. CASTLE adapts ``traditional'' concepts of $k$-anonymity and $l$-diversity to the paradigm of stream processing, enriches them with the possibility to additionally noise the data \cite{castleguard}, and provides flexible adaptations to use case-specific givens (e.g., timeout constraints). Besides, it has been proven to be on par with state-of-the art non-streaming algorithms for $k$-anonymization in matters of privacy metrics~\cite{brunn2023analyzing}. CASTLE has thus been chosen as the starting point for our endeavor. Its underlying clustering approach will be introduced in some more detail in \Cref{sec:rw_bg_castle}.

\subsection{Apache Flink}
Apache Flink is a powerful streaming framework designed for distributed and stateful data processing over continuous streams. Its versatile \textit{Source}, \textit{Process}, and \textit{Sink} architecture supports a wide range of streaming use cases.

The \textit{Source} component defines the input of the data stream, which can originate from systems like Apache Kafka, Cassandra, external APIs, or even static data collections. Once ingested, the data flows into the \textit{Process} phase, where various processing, analytics, and manipulations are performed using defined process functions. These functions operate in a specified sequence to transform the data. After processing, the data exits the framework through the \textit{Sink}, which could involve logging, further streaming via Kafka, storage in Cassandra, or integration with other systems to handle the output data.

Flink is designed for performance and scalability, operating at in-memory speeds and compatible with containerized environments like Docker, making it suitable for diverse production scenarios. Major companies, including Amazon~(Kinesis Data Analytics), Comcast (real-time event stream processing), and Uber~(AthenaX), rely on Flink. Its robust and efficient design solidifies its position as a leading streaming framework, demanding regulatory requirements like data minimization.

\subsection{Data Cluster Anonymization -- CASTLE}\label{sec:rw_bg_castle}
CASTLE~\cite{cao2010castle} anonymizes streaming data in a guarantee-providing manner by clustering incoming data tuples based on their quasi-identifiers until a threshold~$\delta$ is reached, and then anonymizes these clusters through generalization \cite{cao2010castle,p_a_k_anonymity_social_network}.
The parameter $\delta$ sets the maximum delay a data tuple can experience before it must be released. If a tuple reaches a delay of $\delta -1 $, it is published with its corresponding cluster to ensure timely processing.

To achieve $k_s$-anonymity, CASTLE ensures that each cluster contains at least $k$ individuals before generalization. The same applies to $\ell$-diversity, where the diversity of each sensitive attribute in a cluster is checked before generalization.
It is possible that a data tuple belongs to a cluster that has not yet reached the required size of $k$ different individuals when it is about to expire.
In such cases, CASTLE merges this cluster with another to ensure the anonymity constraint while minimizing overall information loss.
Conversely, if a cluster is too large (exceeding $2k$ individuals), it is split into two smaller clusters to reduce information loss while maintaining privacy guarantees.

To preserve as much information as possible--or, vice versa, to minimize the information loss introduced--in this clustering process,
a concept called \emph{enlargement value} is used: Each possible generalization will result in a specific amount of information loss.
When a new data tuple needs to be assigned to a cluster, the additional information loss a tuple would introduce to a given cluster is quantified.
This amount of increased information loss is called enlargement value and is used to determine the optimal cluster for a new data tuple.
The newly arriving tuple is then assigned to the cluster with the smallest enlargement value, balancing privacy and data utility.
Noteworthily, tuples remain in their original form until a cluster is to be released as $k_s$-anonymized output. Only then is the actual generalization applied and the whole cluster is released.

While these mechanisms are suitable for numerical data, CASTLE currently lacks proper loss calculations for non-numerical attributes.
Categorical data, such as `Country' or `Workplace' can only be fully retained (loss $=0$) or completely redacted (loss $=1$), preventing more flexible generalization.
Introducing proper loss metrics (and respective generalization capabilities) for such non-numerical attributes to the pre-existing CASTLE algorithm is therefore one of the core contributions herein.

\section{Proposal: Prink} %
\label{sec:approach}
\begin{figure}[!t]
    \centering
    \includegraphics[width=\linewidth]{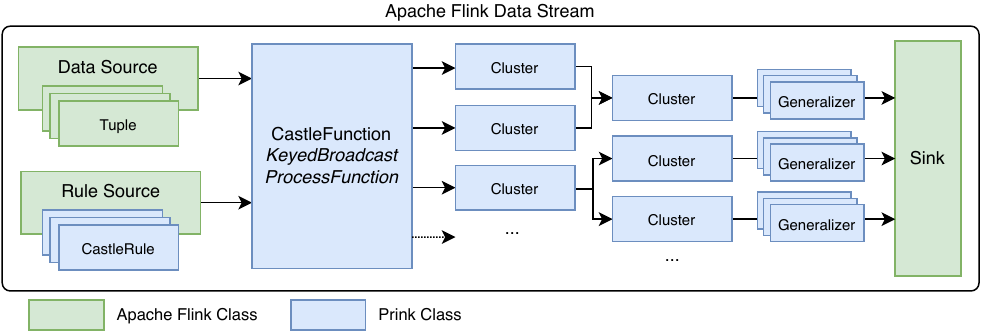}
    \caption{Prink Architecture in the context of Apache flink data streaming infrastructure.}
    \label{fig:architecture_1}
\end{figure}
For our practical implementation called Prink (\underline{Pri}vacy Preserving Fli\underline{nk}), we chose the streaming framework Apache Flink. It offers scalability, fail-safety, and robust stream processing capabilities, making it an ideal choice for integrating anonymization techniques.
Prink was implemented almost entirely within a single Flink \textit{ProcessFunction}, which simplifies integration into existing or new projects.
This approach allows users to easily incorporate Prink at any desired point in their data stream processing pipeline.

Prink’s architecture consists of the following core components, depicted in \Cref{fig:architecture_1}:
\begin{itemize}
	\item \textit{Rule Broadcasting}: Dynamically updates rules for data processing during runtime.
    \item \textit{CastleFunction}: The main processing unit responsible for anonymization and interaction with the Flink data stream.
    \item \textit{Cluster Logic}: Manages groups of data tuples and applies generalization.
\end{itemize}

\subsection{Rule Broadcasting and CastleFunction}
Prink uses Flink's native \textit{Tuple} implementation as input and output formats.
Tuples support flexible structures with up to $25$ attributes of varying data types, making them ideal for dynamic and diverse data streams.
This allows Prink to adapt to missing rules or varying input sizes without requiring significant preprocessing.

Prink achieves dynamic flexibility in data stream anonymization through its integration of \textit{CastleRule} objects and Domain Generalization Hierarchies~(DGH).
The combination of these components ensures that Prink can dynamically adapt to changes in data structures, privacy requirements, and evolving data streams, maintaining both scalability and minimal information loss.

Dynamic adjustments to generalization rules are facilitated by \textit{CastleRule} objects, which are transmitted via a broadcast stream.
Each \textit{CastleRule} specifies how an attribute should be generalized, including configurations for loss metrics and sensitive attribute flags.
This mechanism allows Prink to accommodate runtime changes in data structures and privacy needs, ensuring operational flexibility and adaptability.
To implement these updates efficiently, Prink leverages Apache Flink’s \textit{KeyedBroadcastProcessFunction} rather than a standard \textit{ProcessFunction}.
This specialized function supports two parallel input streams: one for the main data tuples and another for rule updates.
By design, it ensures efficient propagation of runtime rule changes across the system while maintaining the essential keying of data tuples to their respective data subjects.
This keying mechanism is critical for preserving $k_s$-anonymity and $l$-diversity guarantees.
Additionally, the architecture supports parallel execution, allowing Prink to scale efficiently for larger data sets and modern containerized environments.

Unlike static data sets, streaming data often introduces new values or requires updates to the attributes being generalized. Prink addresses these challenges through a twofold approach.
First, DGHs are instantiated at the cluster level rather than being static global structures.
Each CASTLE cluster generates its own DGH instance at creation, ensuring relevance to the current data context.
Outdated hierarchies are replaced as clusters are released, enabling efficient adaptation without the need to re-evaluate existing clusters.

Second, Prink extends DGHs dynamically in real-time. Incoming data tuples can carry hierarchical information within their attributes, such as a value \enquote{Paris} and its corresponding hierarchy ([\enquote{Paris}, \enquote{France}, \enquote{EU}]).
When new relationships are identified, missing nodes and branches are seamlessly added to the DGH of the active cluster.
This allows the hierarchy to evolve alongside the data stream, supporting the dynamic nature of streaming scenarios.
By integrating these capabilities, Prink maintains consistent and contextually relevant hierarchies, even in rapidly changing data environments.

To ensure efficient generalization and minimal information loss, each dynamically created DGH node tracks the number of data tuples it covers.
A node is considered to cover a tuple if its value matches the node or any of its descendants in the hierarchy.
This mechanism enables Prink to calculate optimal generalizations for each cluster, maintaining consistent results while minimizing information loss. By combining the flexibility of dynamic rule updates through \textit{CastleRule} with the adaptability of DGHs, Prink provides a robust and scalable anonymization solution tailored to the complexities of real-time data streams.

\subsection{Cluster Logic}
Within the \textit{CastleFunction}, data tuples are grouped into \textit{Cluster} objects based on generalization rules. 
For that, the \textit{bestSelection} function finds the best cluster for the new data tuple by leveraging the forementioned concept of enlargement value.
Each cluster contains generalizers tailored for different data types, implemented via a \textit{BaseGeneralizer} interface. This modular design simplifies the addition of new generalization strategies, ensuring extensibility for future requirements.
Finally, after $\delta$ tuples were gathered, in the \textit{delayConstraint} function, the first data tuple's cluster gets generalized and published.

\subsubsection{Semantic Generalization Metrics}
\label{sec:generalization}
CASTLE's~\cite{cao2010castle} existing loss metrics for numerical values are insufficient for capturing the complexities of non-numerical, hierarchically generalizable data.
Unlike numerical values, non-numerical attributes cannot be easily ordered or bounded, making traditional numerical approaches inapplicable.
Instead, the calculation of information loss for non-numerical data must be grounded in the structure provided by DGHs, which leverage the inherent categorical and hierarchical relationships between values.

To address this, we adopt a semantic approach to generalization that minimizes information loss while preserving the relationships between non-numerical data points. Simple numerical mappings often fail in this regard, as they disregard the semantic connections between values.
For example, a naive mapping of country/state attributes to numerical values based on their order of appearance loses the semantic relationships, such as the grouping of France and Spain under \enquote{EU} or Florida and Virginia under \enquote{US}.
By using DGHs, we address this issue by grouping non-numerical values into higher-order categories, preserving their semantic structure. To ensure the quality of generalization, we apply information loss metrics tailored to these hierarchies, minimizing disruption to data semantics while maintaining robust anonymization.

The Generalized Loss Metric (GLM), one of the wider used information loss metrics, calculates the information loss by assessing the generalization level of a value within the DGH. Formally, the GLM for non-numerical, i.e., categorical attributes, is defined as:
\begin{equation}\label{eqn:glm_2}
    GLM(u) = \frac{ M_{ p } - 1 }{ M - 1 },
\end{equation}
where $M_p$ represents the number of leaf nodes covered by the current generalization node $u$, and $M$ is the total number of leafs in the hierarchy.
In contrast, for a continuous attribute $a$, the information loss resulting from generalizing its value using an interval $I=[l,u]$ within its domain $[L,U]$ is calculated as $(u-l)/(U-L)$, reflecting the proportion of the domain covered by the generalized interval.

Another metric employed is the Normalized Certainty Penalty (NCP), which evaluates information loss by considering the proportion of values generalized within the hierarchy. For non-numerical data, NCP is calculated as
\begin{equation}\label{eqn:ncp_1}
    NCP_{Cat}(G) = \begin{dcases*} 0, & card(u) = 1 \\ card(u)/\left\lvert A_{ Cat } \right\rvert, & otherwise  \end{dcases*}
\end{equation}
where $card(u)$ is the number of leaf nodes that are present in the sub-tree of node $u$, and
$|A_{Cat}|$ represents the total number of unique categorical values in the data set.
If $card(u)=1$, the information loss is zero, as no generalization is required.
By normalizing the information loss relative to the total attribute diversity, NCP provides a nuanced assessment of the impact of generalization.

Additionally, Prink introduces a dynamic loss metric, the Per Record Request Metric (PRL)~\cite{gadad2019novel,domingo1999resampling},
which adapts to real-time data distributions by factoring in the frequency with which values appear in the stream.
The calculation is based on the Requested Number Count~(RNC), which tracks the number of times a value from the DGH has been requested.
For each generalized node, the total RNC is computed by summing the counts of all its leaf nodes.
This total is then divided by the overall number of requests to the DGH, yielding an information loss value that reflects the relative weight of each leaf node.
More formally:
\begin{equation}\label{eqn:perRecordIL}
    PRL(G) = \begin{dcases*} 0, & card(u) = 1 \\ RNC(u)/RNC(root(u)), & otherwise  \end{dcases*} \text{,}
\end{equation}
where $u$ is the current generalization node.
This metric ensures that frequent values are generalized less, preserving their semantic weight.
The PRL dynamically aligns information loss calculations with the data distribution, making it effective in $k_s$ settings.
To stay adaptable, RNC values are periodically cleared, reflecting evolving weights of attributes.
This is especially valuable in dynamic data flows, where value frequencies fluctuate over time.
By preserving more information for frequent values, the PRL minimizes information loss while maintaining privacy guarantees.

By introducing these three loss metrics to the CASTLE-based anonymization of data streams,
Prink achieves a robust, context-aware approach to generalizing also non-numerical data,
enabling Prink to maintain strong privacy guarantees while minimizing information loss.

\subsubsection{Attribute Weights in Information Loss Calculation}
In Prink, assigning a data tuple to a cluster involves evaluating the information loss for each potential cluster and selecting the one with the lowest average information loss. To provide greater flexibility and align the anonymization process with application-specific priorities, Prink allows the use of attribute weights, implemented as information loss multipliers ranging between $0$ and $1$.

By default, all attributes have equal weight, meaning their contributions to the total information loss are uniform.
In scenarios where certain attributes are more critical for preserving data utility, weights can be adjusted to amplify or reduce their influence.
This adjustment of weights, however, is not required for the functionality of Prink; the system operates effectively even without changing the default weights.

When weights are specified, the information loss for each attribute is scaled by its multiplier. Attributes with larger weights (closer to $1$) have a higher impact on the total information loss, reducing their likelihood of being generalized. Conversely, attributes with smaller weights (closer to $0$) contribute less to the total information loss, making their generalization more likely.

For instance, in cases where attributes are crucial for downstream analysis, assigning higher weights ensures their preservation with minimal generalization. Even if these attributes typically show low information loss under uniform weighting, the applied multipliers prioritize them during the clustering process.

This mechanism ensures that the anonymization process can be tailored to the specific requirements of different use cases. Attribute weights enable the prioritization of certain data characteristics, balancing the trade-off between privacy and utility.
Since adjusting the weights is optional, Prink still provides a fine-grained and flexible approach for a wide range of scenarios, even without modifications to the default weights.

\subsubsection{Design Choices and Extensibility}
To maintain simplicity, all configuration parameters for Prink can be set either through the constructor of the \textit{KeyedBroadcastProcessFunction} or via rule broadcasting during runtime, ensuring minimal user interaction and reducing the risk of misconfiguration.

Prink’s architecture is designed for extensibility. By leveraging the \textit{BaseGeneralizer} interface, developers can add new generalization methods without modifying existing code.
This flexibility is especially valuable for advanced privacy guarantees like $l$-diversity across multiple attributes~\cite{GalCG08}. By ensuring that each sensitive attribute satisfies the $l$-diversity requirement independently, Prink can provide robust privacy protection for data sets containing multiple sensitive dimensions.
Additionally, Flink’s fail-safety mechanisms ensure no data loss during system failures, enhancing the reliability of Prink in production environments.

\section{Evaluation}\label{eval}
In the following we describe our experimental setup including the details of the
data sets and the system environment configurations.
We investigate the impact of our proposal Prink on latency and information loss.

\subsection{Influencing Factors}
\label{sec:influencing_factors}
Before detailing the experimental setup and results, we discuss the factors that influence Prink's performance and information loss to provide a comprehensive understanding. These factors range from the type of streamed data and rule sets used to specific parameter configurations within Prink.

\subsubsection{Parameters}
\label{sec:eval_params}
Prink provides a range of parameters to adjust its functionality to suit specific application needs. Below is an overview of the key parameters and their impact:

\paragraph{Parameter \emph{$k$}:}
Defines the minimum number of distinct individuals required to ensure \emph{$k$}-anonymity, rather than just the number of data tuples. This core parameter directly impacts information loss: as \emph{$k$} increases, more tuples need to be generalized, which can lead to higher information loss. Additionally, \emph{k} interacts with other parameters like \textit{$\delta$} and \textit{$\beta$}, influencing their effects on performance and information loss.

\paragraph{Parameter \emph{$\ell$}:}
Specifies the minimum number of distinct sensitive attribute values required to satisfy \emph{$\ell$}-diversity. Although less influential than \emph{k}, \emph{$\ell$} still plays a significant role in determining both performance and information loss.

\paragraph{Number of sensitive Attributes:}
Prink supports multiple sensitive attributes for \emph{$\ell$}-diversity. An increased number of sensitive attributes requires additional checks, affecting performance. It may also lead to larger generalizations if additional tuples are needed to meet diversity requirements, though this is often dictated by the attribute with the fewest diverse values.

\paragraph{Parameter \textit{$\delta$}:}
Determines how many tuples are retained within Prink before generalization. This parameter significantly affects the number of clusters created and the system's overall performance. A low data flow rate can increase the retention time of tuples, impacting throughput. Since \textit{$\delta$} must always be at least as large as \emph{$k$}, its configuration is critical for maintaining both performance and anonymization guarantees.

\paragraph{Parameter \textit{$\beta$}:}
Limits the number of clusters that can exist simultaneously. While it has minimal impact on information loss, it influences performance by reducing the overhead caused by managing an excessive number of clusters. Balancing this parameter ensures efficient cluster management.

\subsubsection{Generalization Rules}
The choice of generalization rules primarily depends on the data types of quasi-identifiers, such as integers, floats, or categorical values. However, several aspects can be adjusted to influence the results. A key factor is the type of generalizer used, as quasi-identifiers can often be generalized in multiple ways—for example, converting a postal code (integer) to a broader district or suppressing its last digits. The choice of generalizer directly impacts information loss and data clustering, as highlighted in previous sections. Properly matching generalizers to specific quasi-identifiers is thus crucial for effective anonymization.

Another key consideration is the approach used for calculating information loss, which directly impacts generalization and cluster enlargement. Generalization rules specify this method, and their configuration can greatly influence the final outcome.
The Generalized Loss Metric (GLM) is a common default choice and often provides a solid baseline. However, when applied correctly, the Normalized Certainty Penalty (NCP) can produce more optimal results, offering improved information preservation while still meeting anonymization requirements.
For the purpose of this evaluation, we have left the default values unchanged to avoid introducing any fine-tuning that could affect the consistency of the results.

\begin{table}[tb]
    \centering
    \caption{ASHRAE electricity meter readings data set.}
    \label{tab:meter_data}
    \begin{tabularx}{\textwidth}{p{3.3cm}p{1cm}p{5.1cm}r}
    \toprule
    \textbf{Attribute} & \textbf{Type} & \textbf{Range} & \textbf{Unique values} \\
    \midrule
    \verb|building_id| & \verb|int| &  $[565, 655]$ & $89$\\
         \verb|timestamp| & \verb|str| & [2016-01-01 01:00, 2016-12-31 23:00] & $8736$\\
         \verb|meter_reading| & \verb|float| & $[0.0, 2293.88]$ & $124957$\\
         \verb|primary_use| & \verb|str| & Categorical (DGH) & $7$\\
         \verb|square_feet| & \verb|int| & $[387, 420885]$& $89$\\
         \verb|year_built| & \verb|float| & $[1903.0, 2016.0]$& $58$\\
         \verb|floor_count| & \verb|float| & $[1.0, 14.0]$& $13$\\
         \verb|air_temperature| & \verb|float| & $[1.1, 35.0]$& $59$\\
         \verb|cloud_coverage| & \verb|float| & $[0.0, 9.0]$& $4$\\
         \verb|dew_temperature| & \verb|float| & $[-9.4, 17.8]$& $49$\\
         \verb|precip_depth_1_hr| & \verb|float| & $[-1.0, 8.0]$& $4$\\
         \verb|sea_level_pressure| & \verb|float| & $[1007.8, 1031.7]$& $228$\\
         \verb|wind_direction| & \verb|float| & $[0.0, 360.0]$& $37$\\
         \verb|wind_speed| & \verb|float| & $[0.0, 12.9]$& $24$\\
    \bottomrule
    \end{tabularx}
\end{table}

\subsection{Data Set}
\label{sec:data_structures}
To evaluate Prink, we use the publicly available \textit{ASHRAE - Great Energy Predictor III} data set from the Kaggle challenge.\footnote{\url{https://www.kaggle.com/competitions/ashrae-energy-prediction}}
This data set contains meter readings from over one thousand buildings across various sites worldwide.
We chose this dataset specifically because it aligns well with our running example of energy district management, where the goal is to anonymize data at the group level (e.g., neighborhood-level energy usage patterns) while preserving relevant aggregate information.

For our analysis, we merge the training data, building metadata, and weather information into a single data set, focusing exclusively on electricity meter readings and buildings with a given year and floor count.
This preprocessing results in a data set consisting of $321,728$ data tuples representing hourly electricity smart meter readings from $89$ unique buildings, spanning the period from January 1, 2016, to December 31, 2016.
We detail our selected quasi-identifier in \Cref{tab:meter_data}.
Specifically, the \verb|building_id| is used as the unique identifier providing $k_s$-anonymity, while \verb|meter_reading| serves as the sensitive variable.
Although we evaluate only one data set, the use of different quasi-identifiers allows us to cover a variety of data distributions and to
access Prink's capability to generalize different attribute types.
Several factors related to the data structure influence the performance and the resulting information loss.

One critical factor is the number of quasi-identifiers in the data set.
Each additional quasi-identifier increases the complexity of achieving generalization, as all data tuples within a cluster must share the same generalized values for these attributes.
This often leads to greater information loss, despite Prink’s mitigation efforts through its enlargement value concept. Another important aspect is the uniformity of data tuples.
More uniform tuples require less generalization to meet $k_s$-anonymity, reducing information loss.
While this characteristic cannot be controlled in real-world data sets like our evaluation data set, it remains an important consideration in evaluating Prink’s performance.
Lastly, the ratio of individuals to total number of data tuples significantly impacts both information loss and scalability.
Within a cluster, only one tuple per individual contributes to meeting $k_s$-anonymity, so additional tuples from the same individual may require more generalization, increasing information loss.
Furthermore, in parallel execution, Apache Flink ensures that all tuples from a single data subject are processed by the same node, which can limit scalability if the number of data subjects is small.

\begin{figure}[t]
    \centering
    \includegraphics[width=0.7\linewidth]{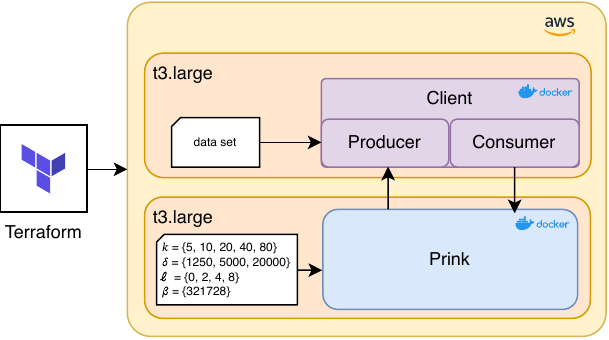}
    \caption{Benchmarking experiment architecture diagram}
    \label{fig:bench-arch}
\end{figure}

\subsection{Experimental Setup}
We now proceed to outline the design considerations, configurations, and automation processes for the benchmarking experiment. An abstract overview is provided in~\Cref{fig:bench-arch}, which will be further detailed in the following section.

The primary objective of the experiment design is to minimize complexity and eliminate uncontrolled interferences wherever possible.
Prink, the system under test (SUT), is deployed on a dedicated virtual machine.
A benchmarking client is used to generate load for the SUT, functioning both as a producer and a consumer. This setup not only simplifies deployment but also closely mirrors real-world architectures, where containerization—most commonly through Docker—is a standard practice.

The benchmarking client, implemented in Go, serves as the load generator for Prink. It operates using a configuration YAML file and an example data set. The YAML file specifies key attributes such as networking ports, addresses, and input/output file paths.
Upon initialization, the benchmarking client loads configuration variables from the config.yaml file along with the evaluation data set.
It begins by launching a single goroutine to sequentially publish each data tuple to Prink.
Each published message includes the exact data tuple values, a unique message ID, and an outgoing timestamp ($t_s$).
Messages are retained in Prink until the specified anonymity guarantees are met, as determined by its configuration and incoming message flow.
Once these conditions are satisfied, the processed messages are returned to the client.
A second goroutine then logs the incoming timestamp ($t_e$) and appends the processed messages to the result log for further analysis.

To evaluate Prink, the SUT, we conducted benchmarking experiments using $60$ distinct parameter configurations, varying key parameters (\emph{k}, \emph{$\delta$}, \emph{$\ell$} and \emph{$\beta$}
as described in \Cref{sec:eval_params} and summarized in \Cref{fig:bench-arch}).
Each configuration was deployed on two AWS \texttt{t3.large} virtual machines (VMs) using an automated Terraform script.
This script handled resource provisioning, Prink configuration, Docker container deployment, and retrieval of results.
To ensure robust evaluation, each configuration was tested three times, totaling $180$ distinct benchmarking runs.
All code and deployment scripts for this evaluation are available in a dedicated GitHub repository\footnote{\url{https://github.com/PrivacyEngineering/prink-benchmark}}. %

\subsection{Evaluation Metrics}
\subsubsection*{Information Loss}
Information loss reduction is one of the core requirements of Prink.
For the metric itself, the evaluation will use the information loss metrics explained in~\Cref{sec:generalization}.
For our evaluation, we use the Generalized Loss Metric for the calculation of the overall information loss.
We calculate the average information loss across all clusters per attribute.

\subsubsection*{Performance}
To ensure Prink's practical viability as an anonymization solution, its performance must be evaluated in terms of latency and its impact on overall application speed. This evaluation focuses on two key aspects of latency.

\begin{figure}[t]
    \centering
    \resizebox{1\textwidth}{!}{\input{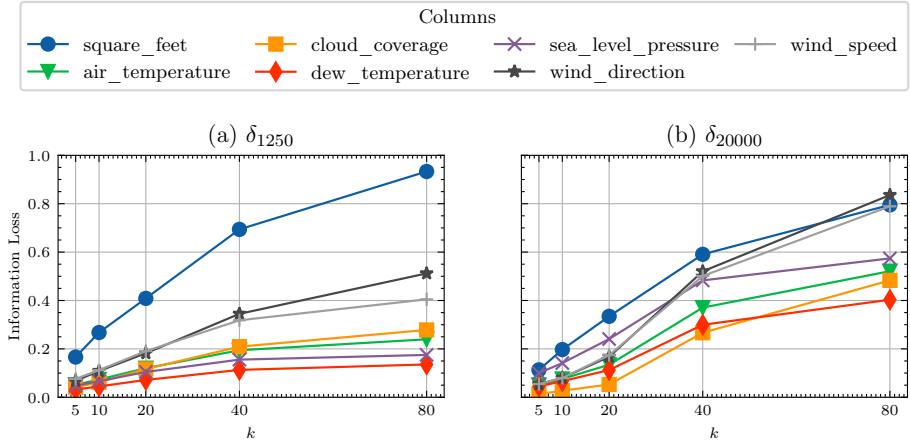}}
    \caption{Evaluation of accuracy through the information loss of attributes.}
    \label{fig:info-loss}
\end{figure}

First, we evaluate the end-to-end latency, which we define as the total time a data tuple spends within Prink before being released back into the main data stream. This latency is determined by calculating as the difference between the timestamp when a tuple enters Prink via the socket and the timestamp when the tuple is returned from Prink through the same socket: $t_e-t_s$.
Notably, since this measurement is performed on the client side, the recorded latency includes network transmission time.

Second, we perform a more detailed analysis of Prink's internal processing times, focusing on key operations. This includes measuring time spent in critical functions such as the \textit{bestSelection} process, where a tuple is assigned to the most suitable cluster based on minimal information loss, and the \textit{delayConstraint} function, which is activated when $\delta$ is exceeded, ensuring timely publication of the cluster containing the oldest data tuple. Furthermore, we measure the \textit{waitTime} between these operations, during which a tuple remains idle, awaiting publication.

Unlike the end-to-end latency measurement, this in-depth analysis is conducted entirely within Prink itself. Consequently, the results are unaffected by network transmission times, offering a clearer and more precise assessment of Prink’s internal processing performance. By differentiating between external and internal contributors to latency, this approach provides a holistic evaluation of the system’s overall efficiency.
\subsection{Results}

\subsubsection{Information Loss}
In \Cref{fig:info-loss}, we present the average information loss for each attribute across different
$k$ values, with $l=1$ held constant,
and note that no attribute weights were specified, meaning all attributes have equal weight.
While it is possible to assign weights to individual attributes, our experiment treats all attributes equally, resulting in generalization being applied uniformly based on the information loss they contribute.

The left-hand \Cref{fig:info-loss} (a) illustrates results for $\delta=1250$, while the right-hand \Cref{fig:info-loss} (b) corresponds to $\delta=20,000$.
An information loss of $0.0$ indicates that no generalization was necessary, and thus no information loss occurred. Conversely, an information loss of  $1.0$ signifies maximum loss, where the attribute was fully generalized.

Regardless of the attribute, the information loss increases as $k$ grows.
This trend is expected, as a higher $k$ requires more distinct individuals---in this case, building IDs with varying attribute values within a cluster--to be generalized.
This pattern is consistent for both $\delta=1250$ and $\delta=20,000$, although the overall information loss is lower for $\delta=20,000$.

The attribute with the highest information loss is \verb|square_feet|, which can be explained by its exceptionally large range ($[387, 420885]$) and relatively few unique values ($89$).
These characteristics often result in extensive generalization into large intervals.
Notably, the ranking of attributes by information loss varies between the two $\delta$ values. For example, \verb|sea_level_pressure| shows minimal information loss at $\delta = 1250$, yet significantly higher loss at $\delta = 20,000$.
This variation underscores how higher $\delta$ values influence generalization strategies differently.
When assigning a data tuple to a cluster, the total information loss across all attributes is calculated, which can lead to discrepancies in attribute-specific generalization. If a particular attribute is more critical for subsequent analysis and requires less generalization, this can be effectively managed by assigning it a higher weight.

\begin{figure}[tb]
    \centering
    \resizebox{.5\textwidth}{!}{
    \input{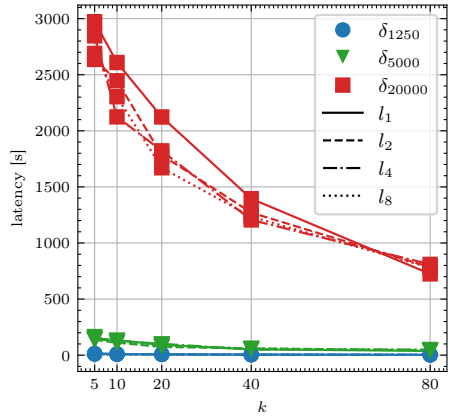}}
    \caption{Evaluation of end-to-end latency, i.e., total time a data tuple spends within~Prink}
    \label{fig:latency}
\end{figure}

\subsubsection{Performance}
In \Cref{fig:latency}, we present the average end-to-end latency,
the total time a data tuple spends in Prink before being returned to the main data stream.
As expected, a higher $\delta$ leads to increased latency, with a
$\delta$ of $1250$ resulting in the lowest latency and a
$\delta$ of $20,000$ producing the highest.
This occurs because $\delta$ directly affects the release timing: with $\delta=1250$, the cluster containing the oldest tuple is released after processing $1250$ data tuples, whereas higher $\delta$ values increase the waiting time before a tuple is published.
Interestingly, latency decreases with higher values of $k$ and $l$,
a trend observed across all parameter settings.
Higher $k$ values result in fewer but larger clusters, which reduces the number of information loss calculations for each arriving tuple.
Additionally, when a cluster is released, more tuples are released simultaneously compared to configurations with smaller clusters.
Together, these effects significantly reduce the overall processing time per tuple.

\begin{figure*}
    \centering
    \resizebox{1\textwidth}{!}{\input{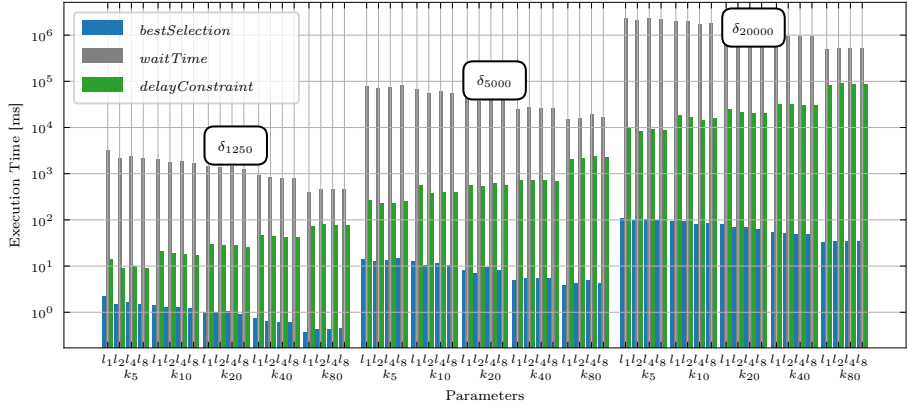}}
    \caption{Evaluation of internal processing times, focusing on key functions: \textit{bestSelection}, \textit{delayConstraint}, and \textit{waitTime} (the interval between these operations).}
    \label{fig:execution-time}
\end{figure*}

Prink's internal processing times are depicted in \Cref{fig:execution-time},
highlighting the time spent in the \textit{bestSelection} and
\textit{delayConstraint} functions, along with the \textit{waitTime} between these two operations.
The x-axis represents various parameter configurations, including three different values for $\delta$, $k$, and $l$.
The y-axis shows the aggregated execution time in milliseconds on a logarithmic scale. Among the three components, the \textit{bestSelection} consistently exhibits the lowest execution time, while the \textit{waitTime} is the longest, highlighting the impact of $\delta$.
As $\delta$ increases, the execution time across all functions rises, a trend that reflects the extended waiting period imposed by higher $\delta$ values.
The \textit{bestSelection} function shows the shortest execution times, decreasing further with larger $k$.
This aligns with the observations in \Cref{fig:latency}, as higher $k$ reduces the number of clusters, thereby minimizing the frequency of information loss calculations.
Conversely, the \textit{delayConstraint} function demonstrates longer execution times as $k$ increases. This is due to the formation of fewer but larger clusters at higher $k$ values, requiring more time for generalization because of the greater number of tuples within each cluster. The execution time is measured as the interval between the initiation of \textit{delayConstraint} and the publication of the final tuple in the cluster. Consequently, the extended processing of these larger clusters leads to a higher average execution time.

\section{Discussion \& Future Work}\label{discussion}
In this section, we discuss the results in terms of the interaction between performance, utility, and privacy, accessing their impact on the approach's applicability and efficiency. We also address privacy considerations and limitations.

\subsubsection*{Trade-off between Anonymity, Utility and Performance}
The results highlight the inherent trade-offs between achieving high levels of anonymity, ensuring low latency, and preserving data utility. In particular, while increasing $k$ provides stronger anonymity guarantees, it also results in higher information loss, as larger clusters require more generalization to ensure indistinguishability among individuals. This interdependency between $k$ and information loss aligns with intuitive expectations and underscores the challenge of maintaining utility as privacy constraints become more stringent.

Our experimental results also reveal an additional dynamic introduced by $k_s$-anonymization: for the same $k$, different $\delta$ values lead to varying levels of information loss and latency. In fact, a higher $\delta$ allows for better generalizations, resulting in lower information loss as clusters can accumulate more tuples over time. However, this comes at the cost of increased latency, as tuples spend more time \enquote{waiting} before being published. Conversely, smaller $\delta$ values prioritize lower latency but may result in higher information loss due to smaller clusters and less effective generalization. This interplay introduces $\delta$ as a new \enquote{tuning parameter}, offering flexibility to balance the competing demands of privacy, utility, and performance in stream-based applications.

\subsubsection*{Practical Feasibility}
The integration of $k_s$-anonymity into a Flink-based stream processing architecture has proven feasible by our implementation, with the system demonstrating the ability to balance these competing goals to some extent. However, achieving optimal performance across all three dimensions—anonymity, latency, and utility—remains a complex challenge, as improving one often comes at the cost of the others. These findings highlight the importance of context-specific configurations and the careful adjustment of parameters such as $k$ and $\delta$ to meet the needs of different use cases.

Future improvements to Prink could enhance both its accessibility and efficiency.
To improve accessibility, Prink could be offered as a packaged Java library via a dependency system, eliminating the need for manual integration. Performance gains could be achieved by caching information loss calculations, reducing redundant computations for unchanged clusters.
Another key challenge is dynamic rule updates, as clusters depend on their initialized rule sets to maintain privacy guarantees, complicating seamless rule changes.

Similar to Gal et al.~\cite{GalCG08},
incorporating weighted sensitive attributes could further minimize information loss while preserving required privacy levels.
In addition, optimal information loss and utility should be investigated across various business domains. Addressing the anonymization of data sets with numerous sensitive features (e.g., in the medical context), and semi- or non-structured data present another challenge.
Extending Prink to support additional privacy-preserving techniques and metrics~\cite{wagner2018technical} further strengthen its effectiveness, particularly in scenarios where adversarial risks are high.

\subsubsection*{Privacy Considerations and Limitations of Prink}
Although $k_s$-anonymity ensures that every released cluster contains at least $k$ distinct individuals, the general limitations of $k$-anonymity still apply. One key issue is that its provided privacy guarantees are not independent of an attacker's background knowledge. If an adversary has external knowledge about the data set, a re-identification of individuals might be possible despite the anonymization.

Another challenge arises from how generalization is applied dynamically.
Since CASTLE adapts its generalization strategies based on the current data structure, the same individual may be generalized differently across time windows. If an attribute remains unchanged but its generalization fluctuates due to temporary data gaps,
it could introduce inconsistencies that might be exploited.

\section{Conclusion}\label{conclusion}
In summary, this paper presents an effective solution for privacy-preserving data anonymization in stream-based applications. We propose adaptations of $k_s$-anonymity tailored for non-numerical streaming data, addressing challenges related to dynamic generalization and privacy guarantees.
Our approach introduces a novel, semantics-aware information loss scheme and supports scalable, distributed anonymization. We also provide a practical implementation, Prink, based on the Apache Flink framework, and demonstrate its performance through a thorough evaluation. By overcoming key limitations of current anonymization techniques, our work enhances the applicability of privacy-preserving methods to real-world scenarios while maintaining strong privacy protections.

\bibliographystyle{splncs04}
\bibliography{references}

\end{document}